\documentclass[preprint,12pt,authoryear]{elsarticle}

\usepackage{natbib}
\setcitestyle{numbers,square}
\usepackage{amssymb}
\usepackage{algorithmicx,algorithm}
\usepackage{makecell}
\usepackage{hyperref}
\usepackage{multirow}
\usepackage{multicol}
\usepackage{stfloats}
\usepackage{graphicx}
\usepackage{color}
\usepackage[justification=centering]{caption}
\usepackage{url}
\usepackage{xcolor}
\usepackage{hyperref}
\usepackage{booktabs}
\usepackage{threeparttable}
\usepackage{appendix}

\journal{Pattern Recognition}

\begin{document}

\begin{frontmatter}



\title{Global Contrast-Masked Autoencoders Are Powerful Pathological Representation Learners}


\author[1]{Hao Quan}

\author[1]{Xingyu Li}

\author[2]{Weixing Chen}

\author[1]{Qun Bai}

\author[1]{Mingchen Zou}

\author[3]{Ruijie Yang}

\author[1]{Tingting Zheng}

\author[4]{Ruiqun Qi}

\author[4]{Xinghua Gao}

\author[1]{Xiaoyu Cui\corref{cor1}}
\cortext[cor1]{Corresponding author:} \ead{cuixy@bmie.neu.edu.cn}

\address[1]{College of Medicine and Biological Information Engineering, 
Northeastern University, China}

\address[2]{Shenzhen College of Advanced Technology, 
University of the Chinese Academy of Sciences, Beijing 100049, China}

\address[3]{School of Computer Science and Engineering, Northeastern University, China}

\address[4]{Department of Dermatology, The First Hospital of China Medical University, China}

\begin{abstract}
Based on digital pathology slice scanning technology, artificial intelligence algorithms represented by deep learning have achieved remarkable results in the field of computational pathology. Compared to other medical images, pathology images are more difficult to annotate, and thus, there is an extreme lack of available datasets for conducting supervised learning to train robust deep learning models. In this paper, we propose a self-supervised learning (SSL) model, the global contrast-masked autoencoder (GCMAE), which can train the encoder to have the ability to represent local-global features of pathological images, also significantly improve the performance of transfer learning across data sets. In this study, the ability of the GCMAE to learn migratable representations was demonstrated through extensive experiments using a total of three different disease-specific hematoxylin and eosin (HE)-stained pathology datasets: Camelyon16, NCTCRC and BreakHis. In addition, this study designed an effective automated pathology diagnosis process based on the GCMAE for clinical applications. The source code of this paper is publicly available at \href{https://github.com/StarUniversus/gcmae}{https://github.com/StarUniversus/gcmae.} 
\end{abstract}



\begin{highlights}
\item We have designed two self-supervised pretext tasks: masking image reconstruction and contrastive learning, which can train the encoder to have the ability to represent local-global features.
\item We discuss the mask ratio, which is suitable for pathology-specific training methodologies based on the masked image modeling paradigm
\item We selected three pathological image datasets and proved the effectiveness of GCMAE algorithm through extensive experiments.
\item An automatic pathology image diagnosis process was designed based on the GCMAE to improve the credibility of the model in clinical applications.
\end{highlights}


\end{frontmatter}


\section{Introduction}
\label{intro}
Pathology is the gold standard of diagnosis. The traditional pathology approach mainly focuses on macroscopic observation under a microscope. Pathologists determine the nature of lesions and classify tissues based on subjective experience, and the diagnosis results are easily affected by factors such as experience and fatigue~\cite{campanella2019clinical,elmore2015diagnostic,wang2016deep}. Computational pathology transforms glass slides into digital images and analyzes them with image processing technology, which promotes the transformation of pathological diagnosis from qualitative analysis to quantitative calculation. In recent years, relying on whole-slide image (WSI) scanning technology~\cite{litjens2017survey}, artificial intelligence algorithms represented by deep learning have achieved remarkable results in the field of computational pathology~\cite{campanella2019clinical,lu2021data,quan2021densecapsnet}; they have become a current research hotspot and a significant direction for the future development of pathology approaches.

Deep learning (DL) is one of the common methods used to extract computational pathology features, as it can directly learn subvisual image features that are difficult for humans to find with their eyes~\cite{pati2021reducing}. However, most DL methods require a large amount of high-quality labeled data, making them difficult to transfer to other datasets with different feature spaces or probability distribution functions~\cite{agarwal2021transfer}. Different staining methods, scanning equipment variations, different diseases and intraclass differences across organs and tissues lead to data feature differences and long-tailed problems, especially in the field of computational pathology~\cite{pati2021reducing}. Maximizing the use of source-domain datasets for representation learning becomes an important method for alleviating the poor model performance caused by data scarcity in the target domain~\cite{agarwal2021transfer,zhuang2015supervised,kolesnikov2020big}.

Recently, self-supervised visual representation learning has achieved great success in the field of natural images~\cite{he2020momentum,chen2020simple,caron2020unsupervised,bao2021beit,chen2020generative,he2022masked}. Self-supervised learning (SSL) can use pretext tasks to mine valuable representation information from large-scale unsupervised data. In the past two years, researchers have applied SimCLR~\cite{chen2020simple}, MoCo~\cite{he2020momentum} and other SSL methods based on contrastive learning to computational pathology and transferred the pretraining models to downstream tasks, thereby narrowing the accuracy gap between unsupervised learning and supervised learning~\cite{dehaene2020self}. In addition, some researchers have designed reasonable data augmentation methods and complex pretext tasks to expand the representation spaces of pathological images according to their characteristics. Yang et al.~\cite{yang2021self} designed a cross-stain prediction and new data augmentation method, stain vector perturbation, based on the characteristics of pathological images and proposed the CS-CO method based on contrastive learning; its effectiveness was verified on NCTCRC datasets. Li et al.~\cite{li2021sslp} developed the SSLP method, which mines pathological features from three aspects, self-invariance, intra-invariance and inter-invariance, by designing complex pretext tasks; this approach surpassed the supervised method on the Camelyon16 dataset. However, the abovementioned self-supervised method based on contrastive learning has the problems of high hardware resource consumption, high training difficulty in multitask learning scenarios, and lower cross-dataset transfer learning performance than supervised learning~\cite{he2022masked,ruder2017overview}. Therefore, simplifying pretext tasks and enhancing the general representation ability of the utilized model are the key problems to be solved in pathological representation learning.

In 2021, as an extensible SSL method, a masked autoencoder (MAE) achieved state-of-the-art (SOTA) results on the ImageNet dataset~\cite{he2022masked}. This method randomly masks part of the input image and uses a lightweight decoder to rebuild the obscured pixels, which can not only yield improved accuracy but also speed up the training process; additionally, its learning efficiency is better than that of contrastive learning~\cite{he2022masked}. Pathological diagnosis often requires the consideration of both the global and local features of WSIs~\cite{campanella2019clinical}. Due to the morphological similarity between cells and tissues of the same type, MAEs may have the potential to find the correlations within pathological image tiles, that is, to extract local features. Correspondingly, if we use the memory bank structure~\cite{wu2018unsupervised} of contrast learning to store the features between each pair of tile, such an MAE may also have the ability to obtain global features.

Based on the above analysis, we propose a global contrast-masked autoencoder (GCMAE)-based SSL model that can extract both global and local features from pathological images. On the one hand, based on the MAE network structure, the model can obtain the internal hidden space feature representation of each patch in pathological images. On the other hand, the model integrates the memory bank structure to store the global features of pathological images, and contrastive learning is used to mine the feature associations between tiles. Second, we also design an automatic pathological image diagnosis process based on the GCMAE for clinical application, which can make full use of unlabeled pathological data to further improve the performance of the model. Finally, we also attempt to utilize a lightweight modeling method to increase the confidence of GCMAE in clinical application. The main contributions of this study are as follows.

1.We have proposed GCMAE, which integrates two self-supervised auxiliary tasks, masking image reconstruction and contrast learning, to produce effective supervision. These tasks also train the encoder to represent local-global features of pathological images.

2.We analyzed the mask ratio suitable for pathological images, and provided guidance for pathology-specific training methods related to the masked image modeling (MIM) paradigm.

3.We selected three pathological image benchmark data sets, and proved that GCMAE has a tangible improvement over other state-of-the-art self-supervised and transfer learning methods through extensive experiments.

4.In this paper, an automatic diagnosis process and a lightweight modeling method for pathological images based on the GCMAE are designed for clinical application purposes.

\section{Related Work}
\label{section:rw}

The paradigm of SSL, which uses the input data themselves to provide supervised signals, has achieved excellent performance in representation learning and has been shown to benefit almost all downstream tasks~\cite{liu2021self}. SSL can therefore make use of the large number of available unlabeled pathology image samples to learn generic representations and migrate them to assist the model learning process in downstream tasks where sample labeling is limited. Currently, the mainstream SSL methods are mainly divided into contrast learning and masked image models, both of which have proven to be effective in pathological image-based visual representation tasks but have shortcomings~\cite{chowdhury2021applying,boyd2021self}.

Contrast learning, as a classic SSL paradigm, has been widely used in the field of visual pathological image representation~\cite{zhang2020contrastive,shurrab2022self}. Several studies have directly applied self-supervised natural image representation algorithms to medical image representation tasks~\cite{dehaene2020self,chen2021momentum}. Although they have been effective, a lack of targeted designs has led to limited performance improvements. A suitable data augmentation approach is one of the key factors required for achieving excellent representation performance with contrast learning~\cite{reed2021selfaugment}. Therefore, some researchers have designed reasonable data augmentation methods for pathological image characteristics, such as elastic deformation~\cite{xu2020data} and stain vector perturbation~\cite{yang2021self}. In addition, specifically designed pretext tasks can also improve the performance of contrast learning in pathological image representation tasks~\cite{li2021sslp}. Some typical works include SSLP~\cite{li2021sslp} and RSP~\cite{srinidhi2022self}, which explore the intrinsic characteristics of pathological images and the multiresolution contextual information that is present with a pyramidal nature, respectively, and these methods have achieved good results. However, two main shortcomings remain in the field of pathological image representation. 1. The homogeneity of the visual appearances of pathological images limits the pathological representation capability of contrast learning; 2. random clipping, as a common data augmentation method that only feeds the sample subject part into the encoder, limits the generic representation capability of contrast learning.

The introduction of MAEs has led to an increasing number of researchers focusing on MIM paradigm~\cite{chen2022context} while addressing the shortcomings of contrast learning for pathological image representation tasks. Recently, several MAE variants have been derived for the task of learning the representations of natural images and videos~\cite{wei2022masked}, but only a few studies have been conducted in the field of medical imaging~\cite{zhou2022self}. One typical approach for pathological images is the SDMAE~\cite{luo2022self}, which, in addition to retaining the original image reconstruction task, adds a self-distillation module to the visible image part for enhanced high-level semantic information learning. However, in a paradigm such as an MAE, which takes image reconstruction as its main task, the encoder can only pay attention to the representation of each sample and lacks the feature relationships between samples, which limits the application of MIM models in the field of pathological representation. Therefore, it is necessary to consider fusing two classic SSL paradigms to design self-supervised models for the characteristics of pathological image representation tasks.

\section{Methodology}
\label{section:mt}
\begin{figure}
\centering
\includegraphics[width=1\linewidth]{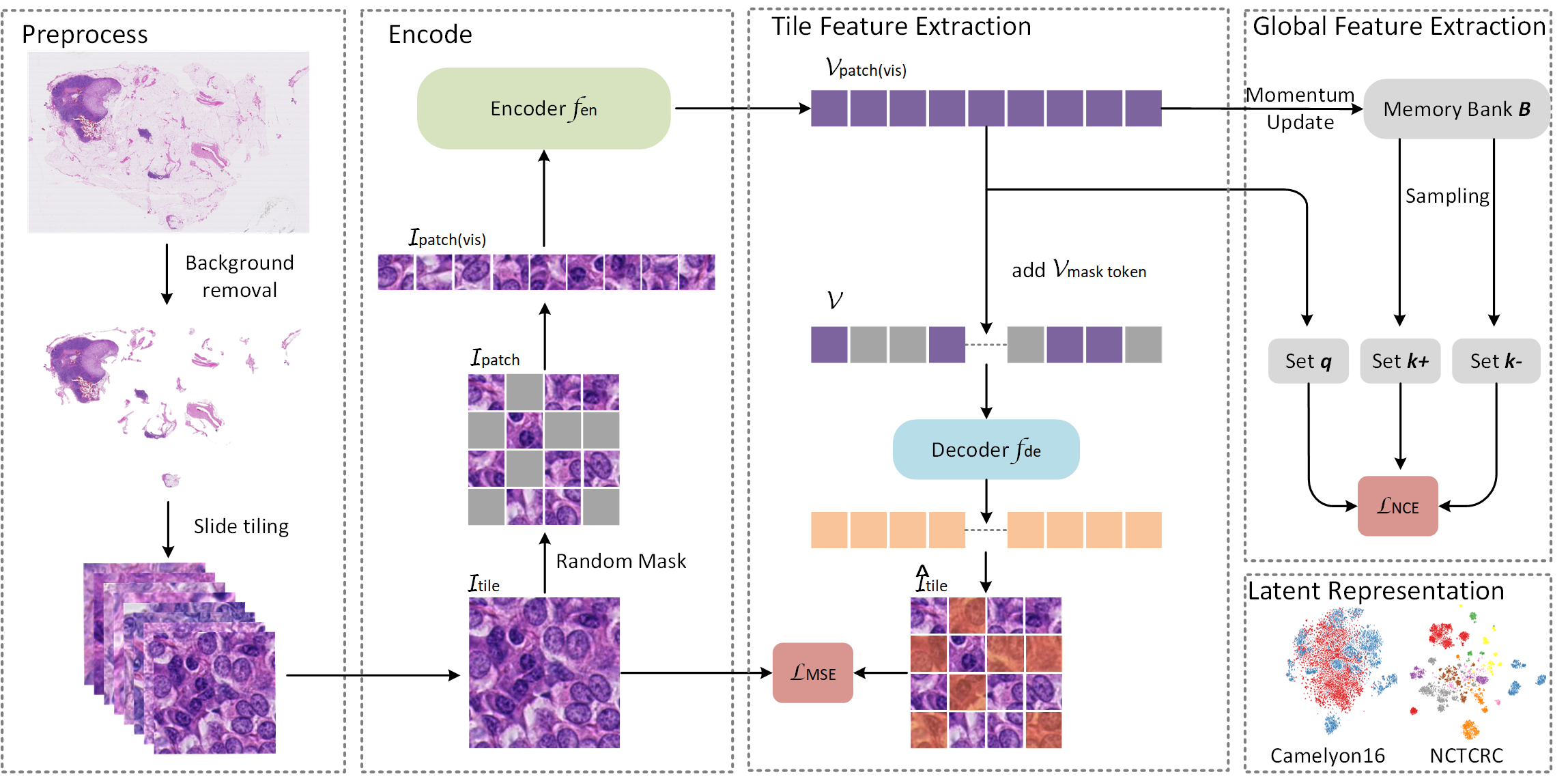}
\caption{\label{fig:figure1}Framework of the GCMAE. Tile feature extraction is the pretext task of image reconstruction, while global feature extraction is the pretext task of contrastive learning. The latent representation is the t-distributed stochastic neighbor embedding (t-SNE) result of the encoder output.}
\end{figure}

In this section, we describe the GCMAE algorithm in detail. Figure~\ref{fig:figure1} shows the framework diagram of our proposed GCMAE-based SSL algorithm. In summary, the GCMAE consists of four parts, a preprocessor, an encoder, a tile feature extractor and a global feature extractor, as well as two pretext tasks: image reconstruction and contrast learning. The GCMAE inherits and optimizes these methods for pathological images. As shown in Eq.~\ref{eq:1}, the weighted sum of the mean squared error (MSE) loss of tile feature extraction and the noise contrastive estimation (NCE) loss of global feature extraction is used as the cost function to reduce the distance between similar features while learning high-level image features, thus improving the generalization of the model and the accuracy achieved in the cross-dataset transfer learning task.

 \begin{equation}
  \mathcal{L} = \lambda_1\mathcal{L}_{MSE} + \lambda_2\mathcal{L}_{NCE}\\
  \label{eq:1}
 \end{equation}

\subsection{Preprocessing} 
The large amount of redundant information contained in WSIs, such as non-tissue background regions, can reduce the training performance of the model. Therefore, it is necessary to perform preprocessing operations on WSIs. First, the optimal segmentation threshold of the given WSIs is calculated based on the Otsu threshold segmentation algorithm, and the tissue regions are extracted. Finally, the mean and standard deviation of the tiles are calculated to achieve the normalization operation, and a normalized image with a mean of 0 and a standard deviation of 1 is output to accelerate the convergence of the model.

\subsection{Encoder} 
The vision transformer (ViT)~\cite{he2016deep} is regarded as the encoder backbone $f_{en}$. Compared with the classic convolutional neural network (CNN), the ViT model without an inductive bias has a very high capacity and a good generalization ability; it can also learn more abundant pathological representations and transfer them to downstream tasks. Because the ViT is a large model, we need to consider an efficient pretraining method to train the visual representation ability of the ViT. An MAE randomly masks some details of the input image with a high mask ratio (MR\%) and reconstructs missing pixels only from the visible part of the feature space, enabling it to achieve excellent performance in natural image representation tasks. This study attempts to extend a simple and efficient MAE to pathological image representation tasks. Specifically, first, 224×224 tiles $\mathcal{I}_{tile}\in{R^{H\times{W}\times{C}}}$ are divided into regular nonoverlapping patches (16×16) $\mathcal{I}_{patch}\in{R^{N\times(p^2\times{C})}}$; a 2D patch sequence is output, where H×W is the size of the original image, C is the number of channels in the image, $p^2$ is the resolution of each patch, and $N=HW/p^2$  is the number of patches cut from the original image. This setting has two main purposes. 1. It is convenient for randomly masking some images. 2. When a 2D image sequence is processed into a 1D image sequence, the sequence length can be reduced. Then, the patches are randomly sampled by a uniform distribution to mask some patches, and the visible parts form a new subset of patches $\mathcal{I}_{patch(vis)}\in{R^{V\times(p^2\times{C})}}$. V=N×(1-MR\%) is the number of visible partial patches and the length of the sequence of valid inputs for the transformer block.

The image features of the visible parts are embedded by linear projection, and the position information is encoded by positional embeddings. Specifically, the 2D image of the visible part $\mathcal{I}_{patch(vis)}\in{R^{V\times(p^2\times{C})}}$ is flatted and mapped to the Dth dimension through a trainable linear projection, and the output vector is a patch embedding. Then, standard learnable 1D vectors are added to the patch embedding to preserve the position information. The embedded features and position information are fed into the transformer block to extract a latent representation of the visible parts of the tile.

\subsection{ Tile feature extraction}
As a decoder $f_{de}$, the tile feature extraction module consists of eight transformer blocks, which form an asymmetric structure with an encoder possessing at least 12 transformer blocks (ViT-base). The asymmetric encoder-decoder structure shows that the encoder and decoder are decoupled, which is beneficial for the encoder to learn more generalized representations. In this study, the decoder mainly assists the encoder in learning general representations. However, while training the encoder, the reconstruction ability of the decoder is also optimized. If a symmetrical structure design is adopted and the encoder and decoder are coupled, even if the encoder's representation ability is insufficient, a powerful decoder can minimize the loss by optimizing the reconstruction ability, thus limiting the feature expression of the encoder. The key to self-supervision lies in pretraining the encoder to attain a strong representation ability and transferring it to downstream tasks. Therefore, it is necessary to adopt an asymmetric encoder-decoder structure, which has also been proven in the field of natural images~\cite{yang2021self}. At the same time, the lightweight decoder design reduces the memory consumption and further expands the application range of the algorithm in clinical environments. The experimental MAE results verify that a decoder with 8 transformer blocks can effectively assist the encoder in learning general representations, and this study follows this setting.

In addition to the latent representation $\mathcal{V}_{patch(vis)}$ , the input also includes a shared, learned vector mask token $\mathcal{V}_{mask-token}$, which is used to indicate the missing patches. The mask token also contains the position embeddings of all patches, which are used to reconstruct the missing pixels. The normalized tiles are used as the target to calculate the MSE loss, as shown in Eq.~\ref{eq:4}.
\begin{equation}
\mathcal{V}=f_{en}(\mathcal{I}_{patch(vis)})+\mathcal{V}_{mask-token}
  \label{eq:2}
\end{equation}
\begin{equation}
\hat{\mathcal{I}}_{tile}=f_{de}(\mathcal{V})
  \label{eq:3}
\end{equation}
\begin{equation}
\mathcal{{L_{MSE}}}=\frac{1}{n}\sum\limits^n\limits_{i=1}(\mathcal{I}_{tile}-\hat{\mathcal{I}}_{tile})^2
  \label{eq:4}
\end{equation}
The random tile sampling strategy can remove redundant information and realize the difficult pretext process of directly reconstructing the original missing pixels from their adjacent patches. However, the strategy can only represent the internal features of each tile and cannot represent the feature relationships between different tiles.

\subsection{ Global feature extraction}
Global feature extraction is implemented through contrastive learning~\cite{kather_jakob_nikolas_2018_1214456}. The latent representation $\mathcal{V}_{patch(vis)}$  is not only inputted to the decoder for image reconstruction but also updated to a memory bank B with a momentum coefficient of t for storing global features. The memory bank is a fixed-length, dynamically updated queue for storing feature embeddings of data samples. In the contrastive learning task, some features can be randomly selected from the memory bank and embedded as negative samples, which can alleviate the limitation of batch size on contrastive learning performance. We design the momentum update feature differently from the momentum update model parameters in MoCo, using only a separate encoder and a memory bank to enable contrastive learning. Specifically, it is known that the latent representation at the output of the encoder in the current epoch is $\mathcal{V}_{patch(vis)}$ , and the latent representation deposited in B from the previous epoch is $\mathcal{V}'_{patch(vis)B}$. The latent representation $\mathcal{V}_{patch(vis)B}$ deposited in B during the current epoch is
\begin{equation}
\mathcal{V}_{patch(vis)B}=\frac{0.5\mathcal{V}_{patch(vis)}+0.5\mathcal{V}'_{patch(vis)B}}{||0.5\mathcal{V}_{patch(vis)}+0.5\mathcal{V}'_{patch(vis)B}||^2}
  \label{eq:5}
\end{equation}
The main reason for this design of $\mathcal{V}_{patch(vis)B}$ is twofold: 1. it can alleviate the differences among the features of the same sample in different epochs caused by the use of different model parameters and random masks during the training process of the network. 2. Combining features from adjacent epochs enables the construction of pathological feature representations with higher information density, thus improving the generic feature representation capability of the GCMAE.

We construct $\mathcal{V}_{patch(vis)B}$ as a dictionary to be stored as a data sample queue in B. The $\mathcal{V}_{patch(vis)B}$ corresponding to the input minibatch of the current epoch is extracted from B as the key value k+. The current minibatch of $\mathcal{V}_{patch(vis)}$  (as query q) forms a positive pair with k. The potential features of the n samples drawn randomly from the memory bank form a negative pair with q. Cosine similarity is used as a means of evaluating the distances between features to calculate the similarity between the $\mathcal{V}_{patch(vis)}$  values. We consider an efficient form of the contrast learning loss function, called infoNCE, to minimize the distances between positive pairs and maximize the distances between negative pairs; the loss function is defined as follows.
\begin{equation}
\mathcal{L}_{q,k^+,B} = -log\frac{exp(q^Tk^+/\tau)}{exp(q^Tk^+/\tau) + \sum_{k^-\in{B}}{exp(q^Tk^-/\tau)}},
  \label{eq:6}
\end{equation}
$\tau$ is a temperature parameter that controls the concentration level of the distribution \cite{boyd2021self}. Because the memory bank dynamically stores a large number of data sample features, the constant comparison and discrimination between q and the memory bank features can help the encoder to effectively mine rich semantic information between tiles and then help the encoder learn global features.

\subsection{Workflow of the proposed GCMAE}
\begin{figure}
\centering
\includegraphics[width=0.8\linewidth]{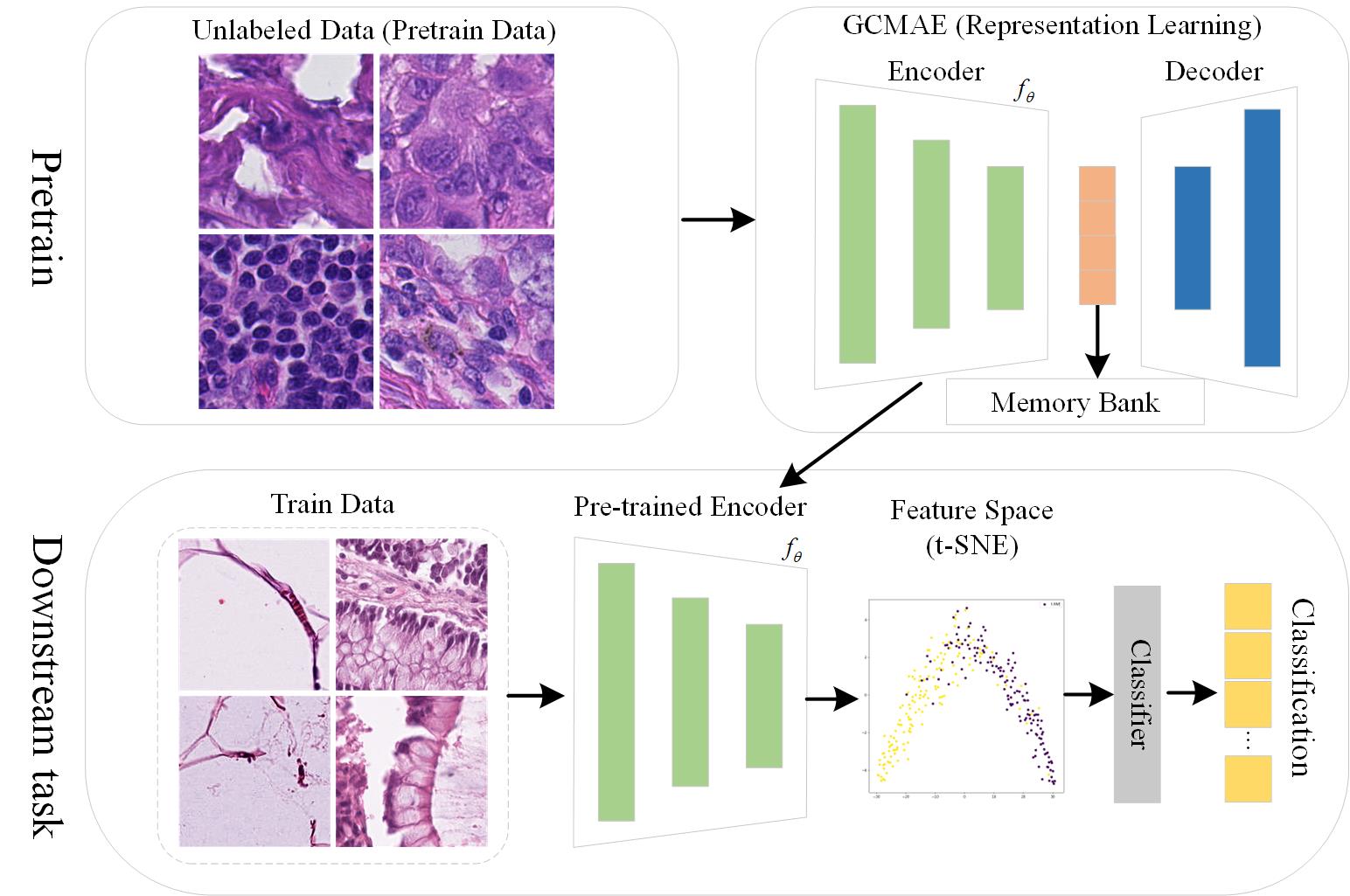}
\caption{\label{fig:figure2}Automatic pathological diagnosis process based on the GCMAE.}
\end{figure}

Based on the GCMAE, we propose a reasonable pathological diagnosis process for images with hematoxylin and eosin (HE) staining. The overall flow chart is shown in Figure~\ref{fig:figure2}, which is mainly divided into three parts: pretraining data collection, the GCMAE and downstream tasks. The pretraining data part mainly prepares a large number of unlabeled pathological image datasets, which are easier to obtain than labeled data. At the same time, the pretraining dataset and the target dataset can be different datasets, which further expands the data sources and reduces the difficulty of data collection. The GCMAE is the key to the whole process, and the encoder is pretrained in a self-supervised way to perform representation learning for pathological images. The good cross-dataset representation ability of the GCMAE-based SSL algorithm improves the classification performance and expansion ability attained in downstream tasks. The downstream task is mainly to fine-tune the pretrained encoder to adapt to the target task by using the target domain dataset. According to different downstream task objectives, the pretraining encoder can be fine-tuned in three ways: 1. taking the pretraining parameters as the initialization parameters of the downstream task model, the best performance can be achieved in the target task by training the model from scratch; 2. the feature extraction part is frozen, only the linear probing method of the classifier is fine-tuned, and the model size and training time are greatly reduced; 3. the model storage weight is reduced from 32 bits to 16 bits by quantization, and the model size is reduced as much as possible while ensuring maximal performance.

\section{Experiment results and analysis}\label{section:ea}

\subsection{Dataset and data settings}

In this paper, three pathological datasets, Camelyon16, NCTCRC and BreakHis, were collected to fully evaluate the general visual pathological image representation ability of the GCMAE-based self-supervised algorithm. The Camelyon16 data set contains two types of breast cancer WSIs. We randomly cut 270k non-overlapping images with the size of 224×224 at 40 magnification. The NCTCRC data set consisted of nine categories of 100k pathological image patches with a size of 224×224, which were scanned at a spatial resolution of 0.5 \(\mu\)m/pixel. BreakHis data set contains 7909 breast tumor pathological images with a size of 700×460 in eight categories. This study mainly uses 400 magnification images in this data set, with a total of 1820 images. Among them, the Camelyon16 and NCTCRC datasets were used for the pretraining and downstream tasks of SSL, and BreakHis was only used as an extended experimental dataset for the downstream tasks. The data setting are shown in Table \ref{table:1}, and the specific data details are shown in the supplementary file part A.

The size of each pathological image used in this study was adjusted to 224 × 224 by the bilinear difference method, and the normalization operation was realized by calculating the mean and standard deviation. In the main comparative experiment of this study, no staining normalization operation was used for pathological images, with the aim of providing harsh training conditions for SSL algorithms to better evaluate their pathological representation abilities and robustness. However, we added the experimental results using stain normalization to further verify the performance of GCMAE.

\begin{table}[htbp!]
\scriptsize
\centering
\begin{threeparttable}
\caption{Data settings for conducting pretraining, training and testing on the three pathological datasets}
\label{table:1}
\begin{tabular}{ccccc}
	\hline
	\multirow{2}{*}{Dataset}	& 	\multirow{2}{*}{Pretrain}		& \multicolumn{2}{c}{Downstream task}	&	\multirow{2}{*}{Overall} \\ \cline{3-4}
			& 			&train	&test						& \\ 	\hline
	Camelyon16	&220,000		&40,000	&10,000	&270,000 \\
 \hline
	NCTCRC	&100,000*		&100,000*	&7,180	&107,180 \\
 \hline
	BreakHis	&N/A			&1,274	&546	&1,820 \\

	\hline
	\end{tabular}
\begin{tablenotes}
\small
\item Note: annotation * is the same data
\end{tablenotes}
\end{threeparttable}
\end{table}

\subsection{Hyperparameter settings and evaluation criteria}
The hyperparameters of the GCMAE method were set as follows: $\tau$ = 0.07, $t$ = 0.5, $k^-$ = 8192, batch size = 128, epochs = 80, and the other hyperparameters were consistent with those of the MAE. With regard to the loss function, we confirmed that $\lambda_1$ = 1 and $\lambda_2$ = 0.1 through pre-experiments to ensure that the losses of MSE and NCE were on the same order of magnitude and that the performance of the GCMAE could be optimized. In order to avoid over-fitting of ViT on relatively small datasets, a reasonable epoch=80 was determined by pre-experiment. The details of the experiment are shown in the supplementary material part B. The optimizer was AdamW with betas= (0.9, 0.95). We set up two classic models as the baselines: ResNet 50 and ViT-base (ViT-B/16). ResNet 50 is a classic CNN and the backbone of MoCo. ViT-B/16 is a high-capacity model with a stronger generalization ability. It is beneficial to build a general representation model in the pathological field when the available pathological data are sufficient, and this model is also the backbone network of the MAE and GCMAE. See part C in the supplementary file for details regarding the hyperparameter settings utilized for the downstream tasks and comparison methods. All self-supervised pretraining experiments did not involve any labels, and all experiments were conducted on an RTX A6000 GPU.

Linear probing and end-to-end fine-tuning are common evaluation methods for self-supervised models. Specifically, linear probing freezes the backbone parameters and trains classifiers in a supervised way. This task focuses on the feature extraction ability of the tested pretraining model and is widely used to evaluate the representation performance of self-supervised models. The end-to-end fine-tuning task involves training a model from scratch on the target task, and the pretraining model is equivalent to the parameter initialization method of the model fine-tuning task. This task is a classic downstream task for a self-supervised model. In practical self-supervised applications, end-to-end fine-tuning can better optimize the target task. This study mainly reports accuracy and Area Under Curve (AUC) to evaluate the model performance. The mean and standard deviation are obtained by running Monte Carlo cross-validation ten times.

\subsection{Mask ratio}
For SSL with the MIM paradigm, images with different information densities are suitable for different mask ratios. Therefore, this study discusses the suitable mask ratio for pathological image representation. Table~\ref{table:2} shows the influence of the mask ratio on the pathological representation of MIM-based SSL, represented by the MAE. In the pathological representation application, a 50\% mask ratio is suitable for linear probing, and an 80\% mask ratio is suitable for fine-tuning, which contrasts with the optimal mask ratio of 75\% in natural image applications. Pathological images contain abundant tissue features, and their information density is higher than that of natural images. Therefore, when using the MAE model to reconstruct pathological images, more information needs to be known. However, the optimal result of model fine-tuning is achieved at a higher mask ratio of 80\%, which shows that the suitable mask ratio for linear probing in the pathological image field is not suitable for model fine-tuning, further proving the result in [18]. Consistent with the natural image results, the influences of pretraining models with different mask ratios on model fine-tuning are lower than that of linear probing, and the accuracy of model fine-tuning is better than that of the ViT-B/16 trained from scratch (81.9\%).

\begin{table}[ht]
\centering
\scriptsize
\caption{Influences of different mask ratios on pathological representation with SSL and the MIM paradigm (\%)}
\label{table:2}
\begin{tabular}{ccccc}
\toprule
\multirow{2}{*}{Mask ratio} & \multicolumn{2}{c}{\textbf{Linear probing}} &  \multicolumn{2}{c}{\textbf{Fine-tunning}} \\
\cline{2-5}
 & Accuracy & AUC & Accuracy & AUC \\
\midrule
10\% & 80.87±0.64 & 89.92±0.84 & 83.41±0.45 & 92.53±0.61 \\
\hline
20\% & 81.79±0.91 & 90.14±0.86 & 83.11±0.79 & 92.07±0.63 \\
\hline
30\% & 82.11±0.88 & 90.75±0.74 & 83.01±0.44 & 91.96±0.31 \\
\hline
40\% & 82.41±0.76 & 90.78±0.69 & 83.24±0.38 & 92.39±0.64 \\
\hline
50\% & \textbf{85.21±0.58} & \textbf{93.18±0.47} & 82.43±0.22 & 92.02±0.33 \\
\hline
60\% & 81.88±0.69 & 90.02±0.74 & 83.22±0.59 & 92.18±0.46 \\
\hline
70\% & 82.25±0.79 & 90.85±0.81 & 83.02±0.37 & 92.33±0.29 \\
\hline
75\% & 81.79±0.84 & 89.82±0.76 & 84.01±0.53 & 92.82±0.49 \\
\hline
80\% & 80.52±0.76 & 88.07±0.59 & \textbf{84.97±0.31} & \textbf{93.52±0.29} \\
\hline
90\% & 80.94±0.73 & 87.14±0.53 & 83.79±0.49 & 92.44±0.36 \\
\bottomrule
\end{tabular}
\end{table}

\subsection{Cross-dataset and cross-disease transfer task}
The lack of a large amount of high-quality labeled data is one of the key factors that limits the performance of deep learning methods. Considering the realistic clinical environment, the long-tailed burst problem has always existed, and it is manifested as follows. 1. A large number of diseases contain only small amounts of medical image data, among which fewer marked data are available. 2. It is difficult to collect high-quality labeled medical image data in the early stages of sudden diseases. Typical cases include COVID-19 and pathological images. A common approach is to use easy-to-collect source domain data to pretrain the model and improve its performance on the target domain data. Therefore, the performance achieved in cross-dataset transfer tasks can better evaluate the value of SSL algorithms.

In this experiment, a difficult cross-dataset and cross-disease transfer task was selected. Specifically, first, the source domain dataset and the target domain dataset were completely different datasets, and they did not intersect (that is, cross-dataset transfer). Second, the source domain dataset and the target domain dataset concerned different diseases (that is, cross-disease transfer). We chose four types of methods for comparison. The first type includes the ResNet50 and ViT-B/16 models, which use random initialization parameters, ImageNet pretraining parameters and pathological dataset pretraining parameters obtained based on supervised classification tasks to complete linear probing and fine-tuning tasks as the baseline of transfer learning. The second type includes three typical methods in the contrastive learning paradigm, SimCLR, MoCo v1 and MoCo v2. The third type includes two typical pathology-specific training methodologies, TransPath \cite{wang2021transpath} and CS-CO. Finally, the fourth type includes a typical MAE algorithm for MIM paradigms. The backbone of contrastive learning paradigm and pathology-specific training method is ResNet 50, and the backbone of MAE and GCMAE is ViT-B/16.

\subsubsection{Transferring from Camelyon16 to NCTCRC}
In this experiment, camelyon16 is used as the pre-training model of the source domain data set, and then transferred to NCTCRC to verify the performance of the model through nine classification tasks. Specific experimental results are shown in Table \ref{table:3}. Experimental results show that the accuracy and AUC of GCMAE are superior to the other four types of methods in linear probing and fine-tuning tasks. Among them, compared with the other four types of methods, GCMAE has significantly improved performance in linear probing tasks, and its accuracy has increased by 10.53\%, 7.47\%, 3.64\% and 3.97\% respectively. Compared with the fully-supervised ViT baseline results (*), GCMAE still has performance improvement, and the accuracy is improved by 7.76\%. This proves that the GCMAE has the ability to learn transferable pathological image representations and has strong generalization. At the same time, linear probing has high training efficiency and less computing power consumption and can quickly transfer to downstream tasks with different targets, which proves that the GCMAE has good scalability. GCMAE still achieves the best performance in fine-tuning tasks, but compared with other SSL algorithms, GCMAE has no significant performance improvement, mainly because the performance of fine-tuning tasks largely depends on the strategy of model training on target data, which weakens the representation performance of SSL algorithm itself.

\begin{table}[ht]
\scriptsize
\centering
\begin{threeparttable}
\caption{Performance comparison among different SSL models when transferring from Camelyon16 to the NCTCRC task (mean±std\%)}
\label{table:3}
\begin{tabular}{ccccc}
\toprule
\multirow{2}{*}{\textbf{Methods}} & \multicolumn{2}{c}{\textbf{Linear probing}} & \multicolumn{2}{c}{\textbf{Fine-tuning}}\\
\cline{2-5}
 & Accuracy & AUC & Accuracy & AUC \\
\midrule
Randomly initailized ResNet 50 & 50.12±1.52 & 63.45±1.56 & 86.75±1.34 & 94.56±1.23 \\
\hline
ImageNet pre-trained ResNet50 & 62.32±0.85 & 75.23±0.98 & 88.12±0.89* & 98.39±0.85 \\
\hline
Camelyon16 pre-trained ResNet50 & 78.69±0.79 & 88.65±0.95 & 89.12±0.96 & 98.42±0.85 \\
\hline
Randomly initailized ViT-B/16 & 43.59±2.32 & 68.89±2.14 & 76.58±2.15 & 84.56±2.06 \\
\hline
ImageNet pre-trained ViT-B/16 & 52.68±1.39 & 88.95±1.62 & 81.46±0.92* & 97.35±0.87 \\
\hline
Camelyon16 pre-trained ViT-B/16 & 73.46±0.85 & 82.23±0.75 & 82.01±0.78 & 96.85±0.84 \\
\hline
SimCLR & 80.95±0.87 & 97.73±0.79 & 90.67±0.52 & 98.99±0.68 \\
\hline
MoCo v1 & 78.40±0.95 & 92.52±0.98 & 89.54±0.45 & 98.57±0.42 \\
\hline
MoCo v2 & 81.75±0.74 & 93.26±0.92 & 91.29±0.85 & 99.02±0.72 \\
\hline
TransPath & 82.41±0.79 & 94.62±0.65 & 91.89±0.46 & 99.05±0.41 \\
\hline
CS-CO & 85.58±0.54 & 98.33±0.41 & 92.01±0.33 & 98.58±0.31 \\
\hline
MAE & 85.25±0.43 & 98.28±0.74 & 93.43±0.47 & 99.12±0.41 \\
\hline
GCMAE & \textbf{89.22±0.32} & \textbf{98.74±0.15} & \textbf{93.89±0.25} & \textbf{99.46±0.19} \\
\bottomrule
\end{tabular} 
\begin{tablenotes}
\small
\item Note: The result of marking * is also the baseline result of fully-supervised learning
\end{tablenotes}
\end{threeparttable}
\end{table}

\subsubsection{Transferring from NCTCRC to Camelyon16}
In this experiment, NCTCRC is used as the pre-training model of source domain data set, and then transferred to camelyon16 to verify the performance of the model through binary classification task, and the specific experimental results are shown in Table \ref{table:4}. Experimental results show that GCMAE is still superior to the other four types of methods in linear probing tasks, and the accuracy is improved by 4.28\%, 1.35\%, 1.71\% and 0.84\% respectively. For fine-tuning tasks, the accuracy of GCMAE is still better than other methods. The AUC of GCMAE is slightly lower than that of MAE, with a performance gap of only 0.16\%.

\begin{table}[ht]
\scriptsize
\centering
\begin{threeparttable}
\caption{Performance comparison among different SSL methods when transferring from NCTCRC to Camelyon16 (mean±std \%)}
\label{table:4}
\begin{tabular}{ccccc}
\toprule
\multirow{2}{*}{\textbf{Methods}} & \multicolumn{2}{c}{\textbf{Linear probing}} & \multicolumn{2}{c}{\textbf{Fine-tuning}} \\
\cline{2-5}
 & Accuracy & AUC & Accuracy & AUC \\
\midrule
Randomly initailized ResNet 50 & 68.72±2.32 & 79.42±2.15 & 80.41±1.75 & 87.62±1.31 \\
\hline
ImageNet pre-trained ResNet50 & 72.96±0.79 & 83.41±0.76 & 82.15±0.49* & 89.68±0.62 \\
\hline
NCTCRC pre-trained ResNet50 & 77.28±0.91 & 85.51±0.76 & 81.12±0.45 & 89.53±0.79 \\
\hline
Randomly initailized ViT-B/16 & 63.51±2.42 & 69.75±2.74 & 78.47±2.21 & 86.22±2.04 \\
\hline
ImageNet pre-trained ViT-B/16 & 69.41±1.25 & 78.27±1.43 & 81.13±0.49* & 88.06±0.52 \\
\hline
NCTCRC pre-trained ViT-B/16 & 75.26±0.85 & 84.02±0.49 & 80.25±0.35 & 87.91±0.51 \\
\hline
SimCLR & 79.29±0.83 & 89.89±0.78 & 80.25±0.85 & 90.69±0.89 \\
\hline
MoCo v1 & 77.82±0.56 & 85.43±0.64 & 80.93±0.45 & 88.75±0.37 \\
\hline
MoCo v2 & 80.21±0.45 & 89.73±0.75 & 81.76±0.43 & 90.87±0.35 \\
\hline
TransPath & 79.85±0.47 & 88.75±0.67 & 82.23±0.51 & 91.45±0.81 \\
\hline
CS-CO & 79.74±0.28 & 88.43±0.32 & 82.47±0.21 & 91.77±0.29 \\
\hline
MAE & 80.72±0.37 & 89.04±0.2 & 83.32±0.15 & \textbf{92.85±0.18} \\
\hline
GCMAE & \textbf{81.56±0.23} & \textbf{90.52±0.32} & \textbf{83.92±0.24} & 92.69±0.16 \\
\bottomrule
\end{tabular}
\begin{tablenotes}
\small
\item Note: The result of marking * is also the baseline result of fully-supervised learning
\end{tablenotes}
\end{threeparttable}
\end{table}

\subsubsection{Performance analysis in cross-dataset and cross-disease transfer tasks}
First, ResNet50 outperforms ViT-B/16 in the transfer learning baseline. We think that the dataset size used in this experiment limits the performance of the ViT-B/16 model. Compared with ResNet, less inductive biases for vision make the ViT have a stronger representational ability, and at the same time, it also leads to ViT being a data-hungry model \cite{ruder2017overview}. This means that the ViT needs more data training models than ResNet to achieve excellent performance. Fortunately, the powerful GCMAE uses only 0.2 million pathological image data for pretraining, which can make the ViT model surpass ResNet on pathological image datasets with small dataset size. This indicates that the low-cost GCMAE has a wider application range in the field of pathological image representation.

Secondly, MAE, a typical model of MIM, is superior to contrastive learning paradigm in pathological image representation task. We believe that the key factors leading to the excellent performance of the MIM self-supervised model are the random mask patch and the reconstruction of missing parts. These settings causes the model to pay attention to the global representation of the input image. In contrast, random cropping, a common data augmentation method used in contrastive learning, can only input the main part of the cropped picture into the network training process, and no other means are taken to urge the encoder to learn the representation for the missing part. As a result, contrastive learning usually pays more attention to the features of the subject part, which inevitably reduces the universality and generalization of the representational transfer process to downstream tasks.

Finally, we also discuss and compare the possible limitations of TransPath, CS-CO and SSLP. The CS-CO training process is divided into two stages: stage one is the generation task of cross-stain prediction, and stage two is a comparison task to fine-tune the encoder obtained by stage one. GCMAE, which is simpler and more efficient, is two synchronized pretext tasks. In TransPath, contrastive learning is implemented through data augmentation of the original image without considering the relationship between tiles. In GCMAE, not only are the features of tiles deeply extracted by generating tasks, but the rich semantics between tiles are fully extracted by comparing tasks. In SSLP, three pretext tasks are successfully designed by analyzing the patch-wise spatial proximity of the WSI. However, for datasets with only tile-level pathological images, the pretext tasks of SSLP may lose their supervisory effect because of the loss of spatial position information. Because SSLP is not open source, it is difficult for us to accurately compare the performance of SSLP, so it is not directly included in the comparison of experimental results. Due to the design of the masking image generation task, GCMAE is superior to the above methods in terms of hardware consumption.

\subsection{Robustness test}

The robustness test mainly evaluates the anti-interference ability of the model to abnormal inputs, which is mainly manifested in the sudden decrease in the dataset size and various noise attacks. Regarding the dataset size, we randomly select 10\%, 50\% and 80\% of the data from the downstream task training set used in the cross-dataset and cross-disease tasks to evaluate the impact of the dataset size on model performance. The main purpose is to observe the changing trend of model accuracy with the decrease of data size. For noise attacks, we choose five kinds of noise to add to the input samples of the model and evaluate the anti-interference ability of the model according to the accuracy of the output results. In this experiment, only linear probing is used to evaluate the performance of the GCMAE. At this time, the weight of the encoder is only obtained through the pretraining of the GCMAE, and no fine-tuning is performed for the target data. The experimental results can better demonstrate the robustness of the pathological representation ability of the GCMAE.

\subsubsection{Influence of the dataset size on the representation ability of the GCMAE}
We mainly compare the performance achieved by the GCMAE with four different dataset sizes. The data setting details for different dataset sizes are shown in Table~\ref{table:5}, in which 100\% denotes the original training dataset; 80\%, 50\% and 10\% are randomly sampled from the original training dataset; and the test set has no change. This experiment is still based on the cross-dataset and cross-disease transfer task. The specific experimental results are shown in Table~\ref{table:6}. The size of the dataset is reduced from 100\% to 10\%, the maximum accessibility difference of ViT-B/16 is 32\%, and the maximum accessibility difference of the GCMAE is only 2.53\%, which is 12.6 times less than that of ViT-B/16. The experimental results show that the dataset size reduction does not significantly reduce the performance of the GCMAE.

\begin{figure}
\centering
\includegraphics[width=0.8\linewidth]{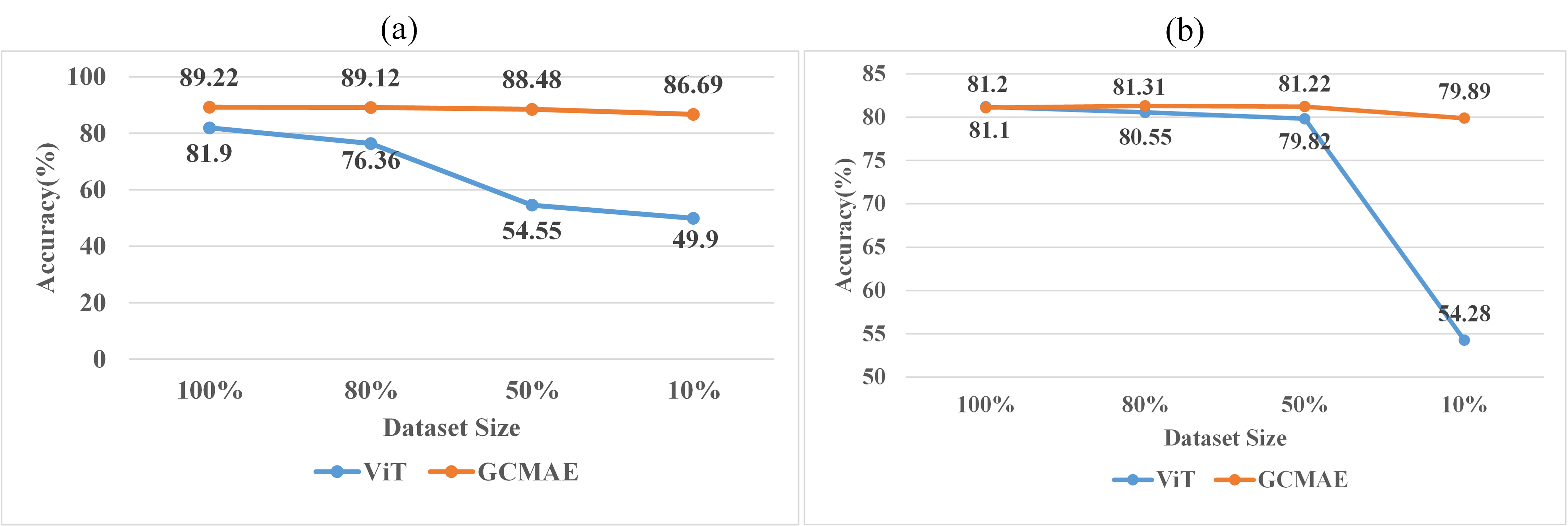}
\caption{\label{fig:figure3}Effects of different dataset sizes on model performance. (A) is the experimental result obtained by pretraining on Camelyon16 and transferring to NCTCRC for downstream tasks, and (b) is the experimental result obtained by pretraining on NCTCRC and transferring to Camelyon16 for downstream tasks.}
\end{figure}

To more intuitively explore the influence of the dataset size on model classification performance, we draw a line chart, as shown in Figure~\ref{fig:figure3}. Compared with the ViT, the GCMAE has more stable performance and almost no accuracy attenuation. In Figure~\ref{fig:figure3}, (a) is a nine-category task and (b) is a two-category task. We find that the more complex the task is, the more significant the performance gain brought by the GCMAE. At present, one of the difficulties in building deep learning models based on medical images is that complex tasks usually require a large amount of high-quality labeled data to achieve good performance. However, the reality is that a large number of diseases only have small amounts of medical imaging data. This limits the application scope of deep learning technology in the field of medical imaging. Our experimental results show that the GCMAE can significantly improve this situation. Pretraining deep learning models based on the GCMAE can significantly reduce the data demand problem faced by large models, which is of great significance for expanding the application scope of deep learning models in the medical imaging field.

\begin{table}[htbp!]
		\scriptsize
		\centering
		\caption{Data settings for different dataset sizes}
	\label{table:5}
	\begin{tabular}{cccccc}
	\hline
	\multirow{2}{*}{Dataset}		&\multicolumn{4}{c}{Train}				&\multirow{2}{*}{Test} \\ \cline{2-5}
							&100\%	&80\%		&50\%		&10\%			& \\ \hline
				Camelyon16		&40,000	&32,000	&20,000 	&4,000		& 10,000\\ \hline
	 			NCTCRC		&100,000	&80,000	&50,000	&10,000		& 7,180\\ 
\hline
\end{tabular}
\end{table}

\begin{table}[ht]
\centering
\caption{Experimental results obtained by the model on tasks with different dataset sizes (mean±std \%)}
\label{table:6}
\resizebox{\textwidth}{!}{%
\begin{tabular}{cccccc}
\toprule
\multirow{2}{*}{\textbf{Task}} & \multirow{2}{*}{\textbf{Method}} & \multicolumn{4}{c}{\textbf{Dataset size}}\\ 
\cline{3-6}
&  & 10\% & 50\% & 80\% & 100\% \\
\midrule
\multirow{2}{*}{Transferring from Camelyon16 to NCTCRC} & ImageNet pre-trained ViT-B/16 & 49.92±1.41 & 54.83±0.98 & 76.53±0.93 & 81.46±0.92 \\
\cline{2-6}
 & GCMAE & \textbf{86.73±0.45} & \textbf{88.52±0.39} & \textbf{89.13±0.31} & \textbf{89.22±0.32} \\
 \hline
\multirow{2}{*}{Transferring from NCTCRC to Camelyon16} & ImageNet pre-trained ViT-B/16 & 54.59±0.89 & 78.93±0.59 & 80.58±0.41 & 81.13±0.49 \\
\cline{2-6}
 & GCMAE & \textbf{79.91±0.43} & \textbf{80.94±0.28} & \textbf{81.39±0.17} & \textbf{81.56±0.23 } \\
\bottomrule
\end{tabular}%
}
\end{table}

\subsubsection{Influence of noise attacks on the representation ability of the GCMAE}

The use of noise attacks is also a common method for verifying the robustness of a model. To test the robustness of the GCMAE, we add Gaussian, salt-and-pepper, uniform, exponential and Rayleigh noises to the test dataset to interfere with the prediction results of the model. Specifically, the initial epsilon value is set to 0.001, the step size is set to a power of 2, and the interval is [0.001, 0.256]. Then, we choose four metrics to evaluate the influence of noise on the performance of the GCMAE model: accessibility, precision, recall and F1 score.

\begin{figure}
\centering
\includegraphics[width=1\linewidth]{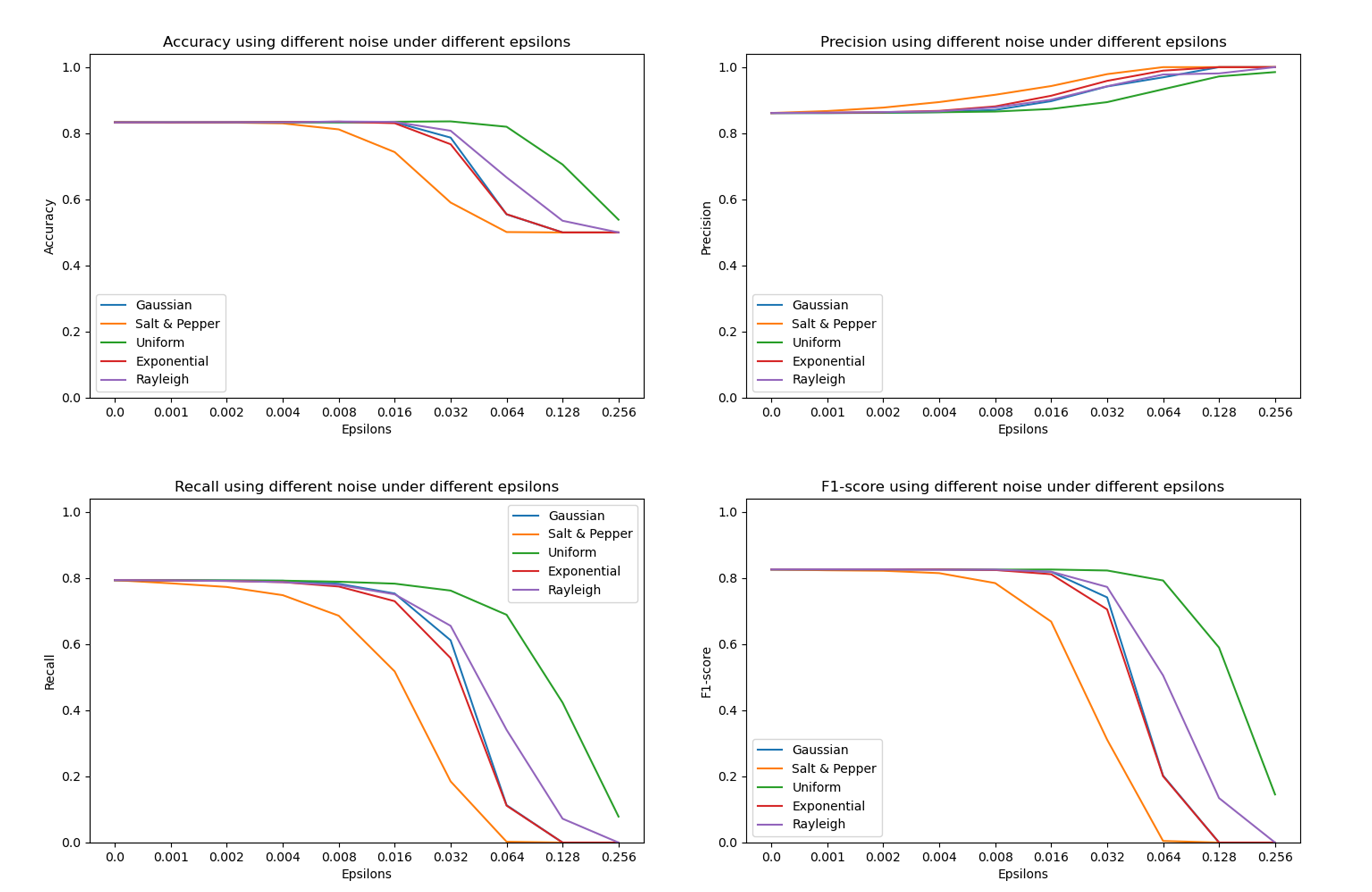}
\caption{\label{fig:figure4}Effects of five noises on GCMAE performance.}
\end{figure}

\begin{figure}
\centering
\includegraphics[width=0.8\linewidth]{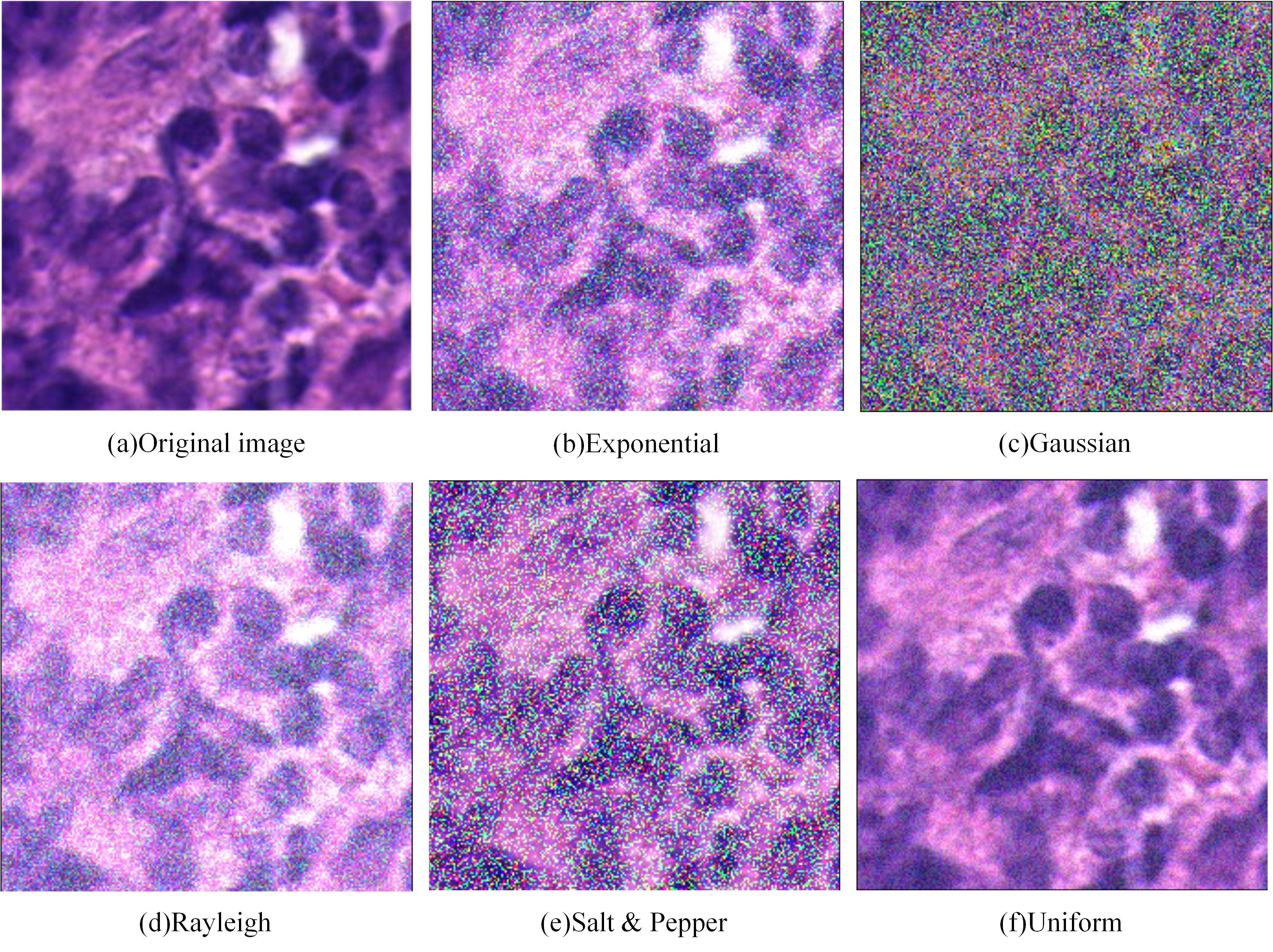}
\caption{\label{fig:figure5}Effects of adding five kinds of noise to pathological images.}
\end{figure}

The experimental results are shown in Figure~\ref{fig:figure4}. First, the GCMAE has the best robustness to uniform noise. When the epsilon value is greater than 0.064, the performance of the model becomes obviously degraded. Second, when the epsilon value is less than 0.016, the GCMAE has good robustness to the other four noises except salt-and-pepper noise, and when the epsilon value is less than 0.004, the GCMAE has good robustness to all five noises. Then, we find that when the epsilon value approaches 0.256, the GCMAE tends to predict tumor images with excessive noise as normal images, which leads to a decrease in the number of false positive samples and an increase in precision. Finally, Figure~\ref{fig:figure5} shows the results obtained after adding noise to pathological images. Direct observation shows that the results of adding uniform noise are closest to the original images, which explains why the GCMAE has the best robustness to uniform noise. Generally, the GCMAE has positive results in terms of suppressing noise interference, and the model has good robustness.

\subsection{Extended experiment}
\subsubsection{Extended experiment on BreakHis}
Breast cancer is one of the most common types of cancer, and its mortality rate is very high. Therefore, we choose the BreakHis dataset to further verify the pathological characterization performance of the GCMAE, and still used the pretrain data of Camelyon16 in Table~\ref{table:1} to pre-train the GCMAE. In this paper, all the pathological datasets used for pretraining the SSL algorithm are collected at a 400× magnification. Therefore, we only use the 400× pathological images in the BreakHis dataset for training and testing with the downstream tasks. 

The specific experimental results are shown in Table~\ref{table:7}, and the GCMAE still achieves the best performance. First, the accuracy of ViT-B/16 on the eight classification tasks of the BreakHis dataset is only 52\%. Training with only 1274 pathological images limits the performance of the ViT, as it is a data-hungry model. This further proves that the size of the dataset has a significant impact on the performance of the model and once again emphasizes that the lack of large-scale available medical image data limits the performance and application scope of a deep learning model. We use the GCMAE algorithm to pretrain the model on the Camelyon16 dataset and transfer it to the BreakHis dataset for downstream tasks. Compared with that obtained when training ViT-B/16 from scratch, the accuracy can be improved by 40.18\% to 92.18\%. The GCMAE algorithm can still achieve good performance on the BreakHis dataset classification task. This proves once again that the GCMAE algorithm has the ability to perform general pathological representation and can be extended to many different diseases. At the same time, the experiment further proves the significance of the GCMAE for improving the performance of deep learning methods in the field of medical imaging.

\begin{table}[ht]
\centering
\caption{Pathological image classification results obtained on the BreakHis dataset (mean±std \%)}
\label{table:7}
\resizebox{\textwidth}{!}{%
\begin{tabular}{ccccc}
\toprule
\multirow{2}{*}{\textbf{Method}} & \multicolumn{2}{c}{\textbf{Linear probing}} & \multicolumn{2}{c}{\textbf{Fine-tuning}} \\
\cline{2-5}
 & Accuracy & AUC & Accuracy & AUC \\
\midrule
Camelyon16 pre-trained ViT-B/16 & 52.42±0.41 & 75.84±0.58 & 51.85±0.37 & 73.22±0.37 \\
\hline
MAE & 53.14±0.34 & \textbf{83.92±0.41} & 89.52±0.27 & 98.51±0.32 \\
\hline
GCMAE & \textbf{53.74±0.26} & 83.79±0.29 & \textbf{92.14±0.17} & \textbf{99.23±0.16} \\
\bottomrule
\end{tabular}%
}
\end{table}

\subsubsection{Influence of Stain Normalization on Transfer Learning Performance}

Staining normalization is a commonly used preprocessing method for pathological images, so we discuss the influence of stain normalization on model performance. From Table \ref{table:3}, we selected five methods with the highest accuracy to compare with GCMAE, which represent transfer learning, contrastive learning, pathology-specific training method and the MIM. The experimental results are shown in Table \ref{table:8}. After using stain normalization, the accuracy of linear probing and fine-tuning of the methods in Table \ref{table:8} has been improved, among which MAE's linear probing has been improved most significantly, and the accuracy has increased by 2.66\%. However, it is still lower than GCMAE without stain normalization. Experimental results show that stain normalization can improve the transfer learning performance of the model and reduce the performance difference between GCMAE and other methods, but it also proves that GCMAE has stronger robustness.

\begin{table}[ht]
\centering
\caption{ Influence of Stain Normalization on Transfer Learning Performance (mean±std \%)}
\label{table:8}
\resizebox{\textwidth}{!}{%
\begin{tabular}{ccccc}
\toprule
\centering \multirow{2}{*}{\textbf{Method}} & \multicolumn{2}{c}{\textbf{Linear probing}} & \multicolumn{2}{c}{\textbf{Fine-tuning}} \\
\cline{2-5}
 & w/o stain normalization & w/ stain normalization & w/o stain normalization & w/ stain normalization \\
\midrule
Camelyon16 pre-trained ResNet50 & 78.69±0.79 & 79.58±0.86 & 89.12±0.96 & 89.33±0.41 \\
\hline
Camelyon16 pre-trained ViT-B/16 & 73.46±0.85 & 74.91±0.74 & 82.01±0.78 & 83.92±0.57 \\
\hline
MoCo v2 & 81.75±0.74 & 82.41±0.62 & 91.29±0.85 & 90.74±0.45 \\
\hline
CS-CO & 85.58±0.54 & 87.56±0.85 & 92.01±0.33 & 93.74±0.51 \\
\hline
MAE & 85.25±0.43 & 87.91±0.47 & 93.43±0.47 & 93.91±0.47 \\
\hline
GCMAE & \textbf{89.22±0.32} & \textbf{90.58±0.49} & \textbf{93.89±0.25} & \textbf{94.21±0.27} \\
\bottomrule
\end{tabular}%
}
\end{table}

\section{Conclusion and Future Work}
\label{section:c}
In this paper, the GCMAE, an SSL method, is proposed. According to the characteristics of pathological images, contrastive learning is introduced to improve the general representation ability of the MAE; this approach achieves SOTA performance in cross-dataset and cross-disease downstream transfer learning tasks, even surpassing supervised learning. This technique is expected to alleviate the influence of the long-tailed medical image problem on the model through cross-dataset pretraining. Second, we discuss the mask ratio that is suitable for pathological representation and provide a reference for a self-supervised algorithm with the MIM paradigm that is applied in the field of pathological image representation. Finally, based on the GCMAE, we propose a reasonable automatic diagnosis process for HE-stained pathological images and explore the feasibility of clinically applying the GCMAE through a lightweight modeling method.

This paper also has some limitations. The main purpose of this work is to construct a general pathological image representation model based on self-supervised learning. Big data and big model are the key factors to build a general representation model. However, in this work, GCMAE only used a single organ, single resolution and a limited number of pathological image data for pre-training. At the same time, due to the limited data size, GCMAE only pre-trains the vit-base model. These deficiencies limit the general representation ability of the model. In the future, we will consider incorporating a large number of pathological image data sets with multi-organs, multi-centers and multi-resolutions, using GCMAE to pre-train large models such as ViT-Large/-Huge, and introducing multi-resolution training ideas, such as Self-path and RSP, to realize encoding multi-resolution contextual information by designing multi-resolution prediction pretext tasks, so as to build a general large-scale representation model of pathological images. At the same time, we will also consider improving the masking strategy to extend the applicability of GCMAE to the Pyramid-based ViTs model.

\section*{Acknowledgements and Conflict of Interest}
This work is supported by the Fundamental Research Funds for the Central Universities N2219001, Ningbo Science and Technology Major Project 2021Z027, Medical and engineering joint fund of Liaoning Province (Grant No.2022-YGJC-76) and National Natural Science Foundation of China (General Program: 82072095). There is no conflict of interest in this paper.



\bibliographystyle{elsarticle-harv} 
\bibliography{references}


\appendix





\section{Data Details}
\begin{figure}[ht]
\centering
\includegraphics[trim={0cm 0cm 0cm 0cm},clip,width= 0.6\textwidth]{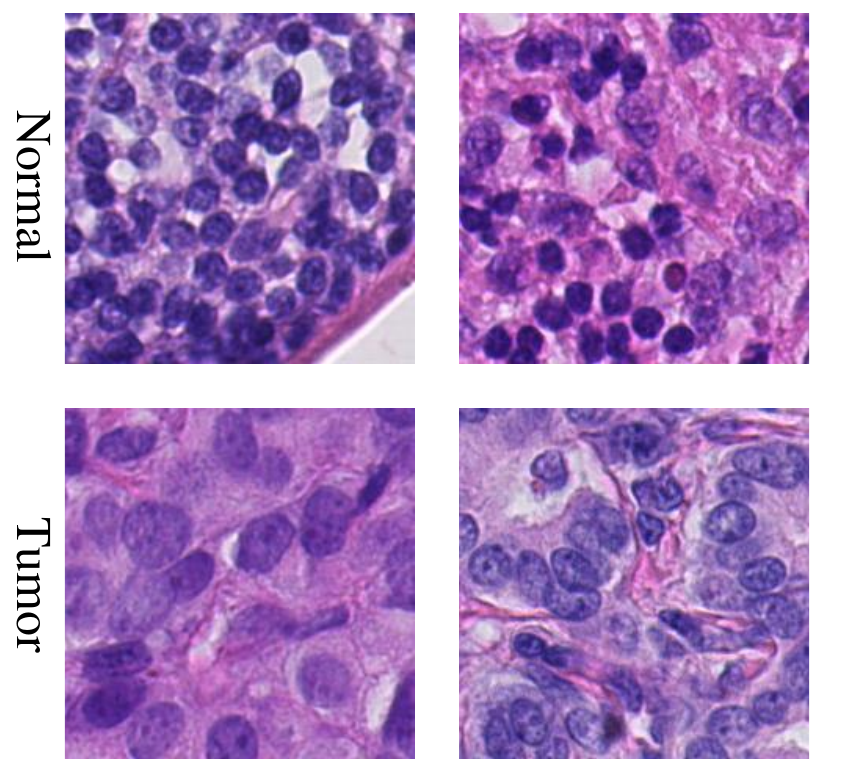}
\caption{Examples of normal and tumor pathological images contained in the Camelyon16 dataset.}
\label{fig:16_dataset}
\end{figure}

Camelyon16 dataset. The dataset contains 400 WSIs of sentinel lymph nodes, among which 70 WSIs containing metastasis have pixel-level labels provided by pathologists. At the same time, the number of pixels contained in the WSIs in this dataset is roughly the same as that in the whole ImageNet dataset, and the amount of data is large. To evaluate the performance differences among different SSL algorithms, we used the complete Camelyon16 dataset to randomly cut 270,000 224×224 nonoverlapping images from normal and tumor tissues at a 40× magnification rate. The ratio of normal tissue to tumor tissue in the dataset used for SSL model training was 10:1, which was set to simulate the data distribution of the Camelyon16 dataset. Some HE-stained pathological images of normal and tumor lymph nodes are shown in Figure~\ref{fig:16_dataset}, and the specific data settings are shown in Table~\ref{table:1}.

\begin{table}[htbp!]
		\footnotesize
		\centering
		\caption{Data settings for conducting pretraining, training and testing on the Camelyon16 dataset.}
	\label{table:1}
	\begin{tabular}{|c|c|c|c|c|}
	\hline
	\multirow{2}{*}{Type of disease} & \multirow{2}{*}{Pretrain} & \multicolumn{2}{|c|}{Downstream task} & \multirow{2}{*}{Overall}\\ \cline{3-4} 
	                                 &                           & train  		& test 		     & \\ \hline
				normal  		 & 200,000   			  & 20,000    		& 5,000   		     & \multirow{2}{*}{270,000} \\   \cline{1-4}  
				tumor 		 & 20,000    			  & 20,000    		& 5,000   		     & \\ \hline 
	\end{tabular}
\end{table}

\begin{figure}[ht]
\centering
\includegraphics[trim={0cm 0cm 0cm 0cm},clip,width= 0.8\textwidth]{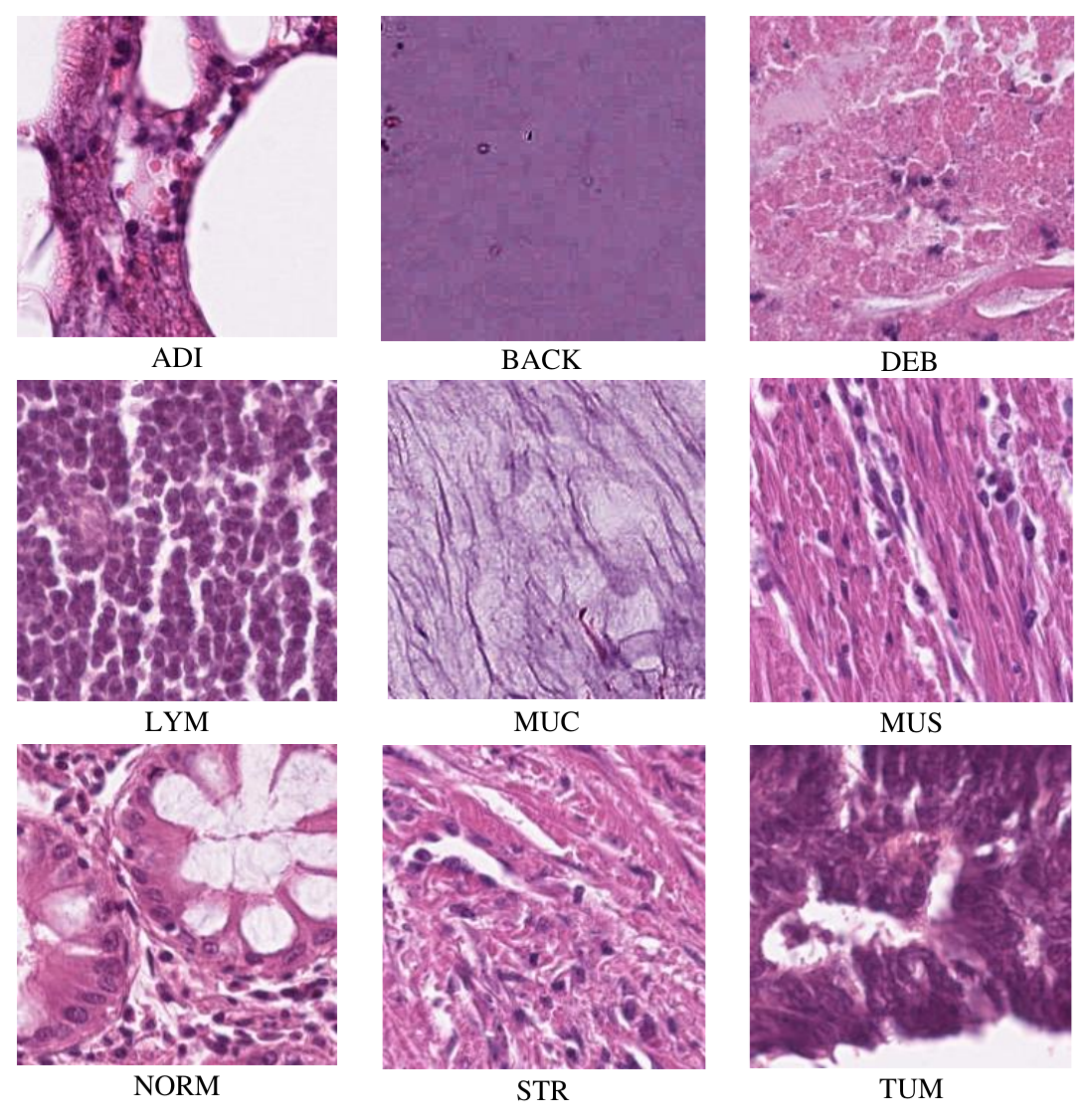}
\caption{Nine types of pathological images contained in the NCTCRC dataset.}
\label{fig:N_dataset}
\end{figure}

NCTCRC dataset. This dataset contains 224 × 224 nonoverlapping images segmented from 86 HE-stained images of human colorectal cancer tissues and normal tissues, which are divided into nine categories, namely, adipose (ADI), background (BACK), debris (DEB), lymphocytes (LYM), mucus (MUC), smooth muscle (MUS), normal colon mucosa (NORM), cancer-associated stroma (STR), and colorectal adenocarcinoma epithelium (TUM). Among them, the training set contains 100,000 images, and the test set contains 7,180 images. In this study, the complete NCTCRC dataset was used to evaluate the performance of the proposed model, where the training set was used for the pretraining of the self-supervised model and the training process employed in downstream tasks, and the test set was used for downstream task testing. Images of the nine types of tissues in the NCTCRC dataset are shown in Figure~\ref{fig:N_dataset}, and the specific data settings are shown in Table~\ref{table:2}.

\begin{table}[htbp!]
		\footnotesize
		\centering
		\caption{Data settings for conducting pretraining, training and testing on the NCTCRC dataset.}
	\label{table:2}
	\begin{tabular}{|c|c|c|c|}
	\hline
	Type of disease	& Pretrain/Train		& Test	& Overall \\ \hline
		ADI		&	10,407			& 1,338	& \multirow{9}{*}{107,180} \\ \cline{1-3}
		BACK		&	10,566			& 847		& \\ \cline{1-3}
		DEB		&	11,512			& 339		& \\ \cline{1-3}
		LYM		&	11,557			& 634		& \\ \cline{1-3}
		MUC		&	8,896				& 1,035	& \\ \cline{1-3}
		MUS		&	13,536			& 592		& \\ \cline{1-3}
		NORM		&	8,763				& 741		& \\ \cline{1-3}
		STR		&	10,466			& 421		& \\ \cline{1-3}
		TUM		&	14,317			& 1,233	& \\ \cline{1-3}
		\hline

	\end{tabular}
\end{table}

\begin{figure}[ht]
\centering
\includegraphics[trim={0cm 0cm 0cm 0cm},clip,width= 1\textwidth]{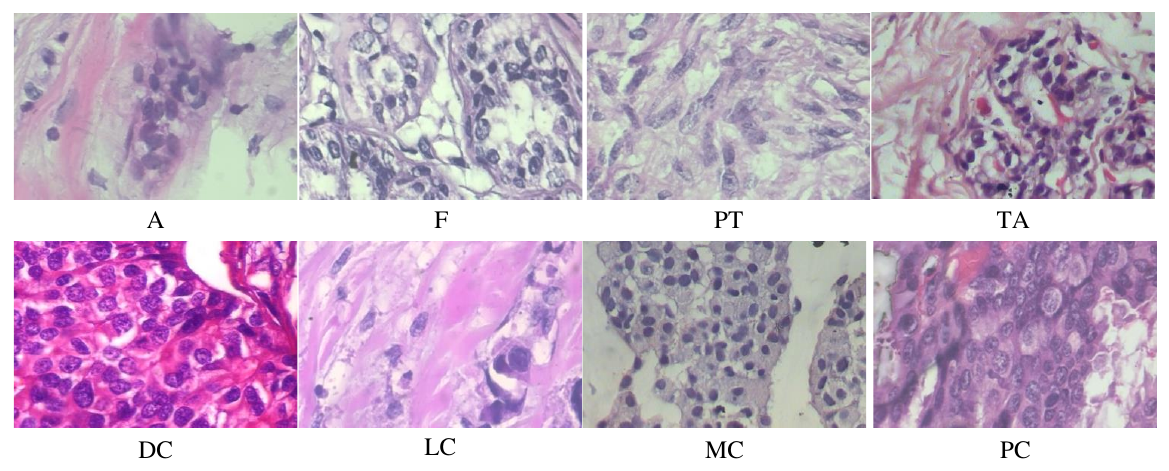}
\caption{Examples of eight types of pathological images contained in the BreakHis dataset.}
\label{fig:B_dataset}
\end{figure}

BreakHis dataset. This dataset is a database of breast cancer histopathological images, which include HE-stained microscopic images with different magnifications. The dataset contains 9,109 microscopic images of benign and malignant breast tumors, which are divided into four benign categories, adenosis (A), fibroadenoma (F), phyllodes tumor (PT), and tubular adenona (TA), and four malignant categories: carcinoma (DC), lobular carcinoma (LC), mucinous carcinoma (MC) and papillary carcinoma (PC). According to the classifications of benign and malignant lesions, the data contain 2,480 benign samples and 5,429 malignant samples, and the image size is 700 × 460. To ensure the same magnification as that in the above dataset, only 400× BreakHis data were used in this study, including 588 benign samples and 1,232 malignant samples (totaling 1,820). As an extended experimental dataset, this dataset was only used for training and testing the downstream tasks of the self-supervised model to evaluate the model performance. The eight types of tissue images in the BreakHis dataset are shown in Figure~\ref{fig:B_dataset}, and the specific data settings are shown in Table~\ref{table:3}.

\begin{table}[htbp!]
		\footnotesize
		\centering
		\caption{Data settings of the BreakHis dataset for training and testing.}
	\label{table:3}
	\begin{tabular}{|c|c|c|c|}
	\hline
	Type of disease	& 	Train		& Test	& Overall \\ \hline
		A		&	74		& 32		& \multirow{8}{*}{1,820} \\ \cline{1-3}
		F 		&	166		& 71		& \\ \cline{1-3}
		PT		&	80		& 35		& \\ \cline{1-3}
		TA		&	91		& 39		& \\ \cline{1-3}
		DC		&	552		& 236		& \\ \cline{1-3}
		LC		&	96		& 41		& \\ \cline{1-3}
		MC		&	118		& 51		& \\ \cline{1-3}
		PC		&	97		& 41		& \\ \cline{1-3}
		\hline

	\end{tabular}
\end{table}

\section{SSL pre-experiment}
The dataset used in this study is relatively small, and ViT, as a large model, is prone to overfitting. Therefore, this study designed a self-supervised pre-experiment to determine the appropriate number of epochs for mitigating the impact of overfitting on the representation ability of the SSL model as much as possible. Specifically, this study used the self-supervised MAE algorithm to pretrain the ViT-B/16 model on the data shown in Table~\ref{table:1} and set epochs = 120. Then, the model saved at epochs=40/60/80/100/120 was migrated to the downstream task of linear probing to evaluate the performance of the pretraining model. The specific experimental results are shown in Table~\ref{table:4}. The results showed that epochs=80 was suitable for the current data scale and achieved the best performance in terms of two indices. Therefore, all SSL experiments in this paper followed this setting.

\begin{table}[htbp!]
		\footnotesize
		\centering
		\caption{SSL pre-experiment.(\%)}
	\label{table:4}
	\begin{tabular}{|c|c|c|c|c|c|}
	\hline
	Epochs	&40		&60		&\textbf{80}		&100		&120\\ \hline
	Acc		&80.08	&80.63	&\textbf{81.33}		&80.22	&78.72\\ \hline
	AUC		&88.79	&89		&\textbf{89.08}		&88.95	&87.27\\

	\hline
	\end{tabular}
\end{table}

\section{Hyperparameter Settings}
\begin{table}[htbp!]
		\footnotesize
		\centering
		\caption{Main hyperparameter settings for different tasks.}
	\label{table:5}
	\begin{tabular}{|c|c|c|c|c|c|}
	\hline
							& Backbone		& Batch size	& Epochs	& Warmup epochs	 & norm pix loss  \\ \hline
	\multirow{2}{*}{Baseline}		& ResNet50		& 128			& 90	     & N/A		      & N/A \\ \cline{2-6}
							& ViT-B/16		& 128			& 90		& 10			 & N/A \\ \hline
				MoCo v1		& ResNet50		& 128			& 80		& N/A 	     	 & N/A \\ \hline
				MoCo v2		& ResNet50		& 128			& 80		& N/A      	 & N/A \\ \hline
				MAE			& ViT-B/16		& 128			& 80		& 5			 & TRUE \\ \hline
				GCMAE			& ViT-B/16		& 128			& 80		& 40			 & FALSE \\ \hline
				Linear probing	& ViT-B/16		& 512			& 90		& 10			 & N/A \\ \hline
				Fine-tuning	& ViT-B/16		& 128			& 50		& 5			 & N/A \\
	\hline
	\end{tabular}
\end{table}

\section{Lightweight modeling}

Considering the actual situation in clinical applications, we discuss the feasibility of the lightweight GCMAE-based pretraining model. Specifically, the pretraining model is compressed by a quantitative method.

A common lightweight method is to compress the employed pretraining model by quantization technology. Specifically, we compress the weight values stored in the GCMAE pretraining model from 32 bits to 16 bits via quantization technology and compress the model size as much as possible on the premise of ensuring accuracy. The performance of the lightweight GCMAE is shown in Table \ref{table:6}. The training time required for completing an epoch on the Camelyon16 pretraining dataset shown in Table \ref{table:6} decreases from 4.06 min to 3.13 min, and the time is shortened by 22.9\%. The time required to complete the test with 10000 images decreases from 0.23 min to 0.18 min, and the time is shortened by 21.7\%. It is predicted that one image takes only 1 millisecond. The parameter size of the model decreases from 982.12 MB to 421.58 MB, and the size is reduced by 57.1\%. Regarding the classification performance of the model, the performance degradation induced by the lightweight GCMAE is only approximately 2\%.

Generally, the above experiments prove that the model size and training time of the GCMAE pretraining model can be successfully compressed by some lightweight methods on the premise of ensuring classification performance. This is of great significance for expanding the clinical application scope of the GCMAE and enhancing the confidence in its clinical feasibility.

\begin{table}[htbp!]
\footnotesize
\centering
\caption{Time consumption levels, parameter sizes, and performance of the lightweight model (\%)}
\label{table:6}
\begin{tabular}{|c|c|c|c|c|c|}
\hline
Model			&Training time	&Test time	&Parameter size		&Acc		&AUC \\ \hline
GCMAE				&4.06 min	&0.23 min	&982.12 MB			&83.92±0.24		&92.69±0.16 \\ \hline
Lightweight GCMAE	&3.13 min	&0.18 min	&421.58 MB			&82.31±0.75		&91.58±0.64 \\ \hline
\end{tabular}
\end{table}

\end{document}